\documentstyle[epsfig,12pt]{article}

\begin{document}

\title{ A Merton-Like Approach to Pricing Debt based on a non-Gaussian  Asset Model }

\author{Lisa Borland, Jeremy Evnine  and  Benoit Pochart\\Evnine-Vaughan Associates, Inc.\\
456 Montgomery Street, Suite 800 \\ 
San Francisco, CA 94133,  USA\\ 
E-mail: lisa@evafunds.com}
\maketitle

\abstract{
This paper is a contribution to the  Proceedings of  the Workshop
{\it Complexity, Metastability and Nonextensivity } held in Erice 20-26 July 2004, to be published by World Scientific. We propose a generalization to Merton's model for evaluating credit spreads. In his original work, a company's assets were assumed to follow a log-normal process. We introduce fat tails and skew into this model, along the same lines as in the option pricing model of
Borland and Bouchaud (2004, Quantitative Finance {\bf 4}) and illustrate the effects of each component. Preliminary empirical results indicate that this model fits well to empirically observed credit spreads with a parameterization that also matched observed stock return distributions and option prices.
}

\section{Introduction}

Recent years have brought with them a vastly growing interest in credit markets and the related credit derivatives markets. The amounts trading on these markets are  growing rapidly.  As the liquidity has increased, so has the interest in models which can price credit risk. Indeed, one of the most esteemed pricing models was proposed already in 1974 by Merton \cite{merton}, although it is only recently  that the credit markets have become liquid enough and transparent enough to actually use these models to price the yield on risky corporate bonds, whose spread to Treasuries is also called the credit spread. It is quite interesting to note, as 
  Miller and Modigliani pointed out in 1958 \cite{miller},  that there is a relationship between a firm's debt and its equity in that the sum of the two are equal to the total assets of the company.  Therefore, a company should be indifferent to how it finances its projects, via debt or equity.
On the other hand, the equity markets and bond markets are so much more developed than the  credit and credit derivatives markets. 
This discrepancy together with the intimate relationship between the different markets implies  that there could be arbitrage opportunities if relative mispricings exist between the fair value of the debt relative to the fair value of the equity. Such a
capital structure arbitrage has been a very popular trading strategy, especially in the early 2000's.

In reality, the capital structure of a company is quite complex.  For example, corporate debt has different levels of seniority, which defines the order in which the debt holders get paid. Clearly, corporate debt is risky since there is a chance that the company will default before being able to pay back its debt.
Equity holders get paid after all debt has been repaid, but even among stock holders there are different priority levels. 
However, as Merton pointed out \cite{merton},
in the simple case where a company has just one type of debt outstanding, the  stock price can  be seen as a call option on the  total underlying assets, struck at the face value of the debt.  Likewise, the corporate bond can be interpreted as long a riskless bound plus a short put on the assets, struck at the face value of the debt. 
Therefore, using the Black, Scholes and Merton option pricing theory \cite{blackscholes,merton1}, it is possible to price corporate debt with the techniques of option pricing.  Merton's structural model, and similar models which derive from Merton's ideas, constitute a widely used framework for pricing debt and credit risk, and for obtaining estimates of the risk-neutral probability of default of a  company.

The basic strength of Merton's model lies in the ability to price the debt as an option on the underlying asset process, via standard tools of option pricing theory. The approach is complicated by the fact that in reality, the asset process is not observable and must be backed out from the stock process.
It is typical that a log-normal distribution is assumed for the asset returns,
  and that the driving noise for both the asset and the stock dynamics follows a 
 standard Brownian process.

 However, it is quite well-known that the empirical distribution of log returns is highly non-Gaussian, exhibiting fat tails which are neglected in the standard Black, Scholes and Merton option pricing theory. Indeed, in options markets one observes that the standard theory underestimates the prices of options at strikes below and above the current stock prices. This means that one must use a higher volatility parameter in conjunction with the Black-Scholes-Merton theory in order to correctly price the options. A plot of this volatility versus the strike price generally forms a  concave function, rather than a straight line which would be the case if the model was "perfect". There have been several attempts in the literature to accommodate this fact, including
stochastic volatility models,  multifractal models, local volatility models, models that
assume fat-tailed random noise such as Levy noise, GARCH-like models and recent multi-timescale models (for an up-to-date review see \cite{mandel}). However, these models are often quite complex and the simplicity of the Black-Scholes-Merton approach is lost. A unique martingale measure is typically not found, neither are closed-form solutions.

 An alternative approach which one of us proposed recently relies on modeling the noise as a statistical feedback process \cite{prl_borland,qf_borland,qf_borlandbouchaud} which yields non-Gaussian distributions for the stock returns yet maintains many useful features of the
standard theory.
In particular, closed form solutions which generalize the Black-Scholes-Merton model are found. These incorporate both fat tails and skew, two features of real returns which are absent in the standard theory.

Since the non-Gaussian theory has shown some success in being able to price options on the underlying equity in a parsimonious fashion which matches well to empirical observations, our goal in this paper is to explore whether the same can be said about pricing credit. The basic notion is to extend Merton's model for pricing credit risk into the non-Gaussian framework. We shall then explore the effects of introducing fat tails and skew into the model. Finally we shall report some preliminary empirical results which indeed lend support to our approach.

\section{Merton's model}

We proceed with a brief review of Merton's model.
He developed a structural model relating the equity and debt markets by thinking of both stocks and bonds as being contingent claims on the same underlying, namely the assets of the company.  His model assumed the assets $A$ follow a standard log-normal process,
\begin{equation}
dA = \mu A dt+\sigma_A A d\omega
\end{equation} where $\omega$ is a standard Brownian noise, $\delta$- correlated in time. 
As follows from the theorem of  Miller and Modigliani, the value of the firm in invariant to its capital structure or in other words, that $A$ is equal to the sum of its debt $D$ and stock (equity) $S$, such that the only relevant variable is the leverage $L= D/A$ . Furthermore, one assumes that it is possible to continuously trade the assets. (Note that this is unrealistic – in reality
 the stock and the bonds are traded, but not the assets themselves.) The interest rate $r$ is assumed constant over time. 

In Merton's model, the scenario is such that the company has issued bonds  of face-value $D$ that will become due at a future time $T$. These bonds are assumed to be zero-coupon which means that there are no intermediate payments before expiry. In addition,
the company has issued equity with no dividends payments. If at time $T$, the value of the assets $A$ is less than the promised debt payment $D$, then the company defaults. The debt holders receive $A < D$ and the equity holders receive nothing. If instead $A>D$, the debt holders get $D$ and the stock holders get $A-D$. The equity of the company can therefore be seen as a European call option on the assets with maturity $T$ and a strike price equal to $D$.
In the standard framework this is just given by the closed-form Black-Scholes-Merton call option pricing formula (cf \cite{hull}), namely
\begin{equation}
\label{eq:scall}
S_0 = A_0 N(d_1) \, - \, D e^{-rT} N(d_2)
\end{equation}
where
\begin{eqnarray} 
d_1 &= &\frac{\ln(A_0e^{-rT}/D)}{\sigma_A \sqrt{T}} + 0.5 \sigma_A\sqrt{T}\\
d_2 &=& d_1-0.5 \sigma_A \sqrt{T}
\end{eqnarray}  and $N(d)$ is the cumulative normal 
distribution calculated up to the limit $d$. Note that $ \tilde{D} = De^{-rT}$   is the
 present value of the promised debt  payment, so this expression  can be expressed in terms of the leverage $L = \tilde{D} /A_0$, namely
\begin{equation}
S_0 = A_0 (N(d_1)  \, - \, L N(d_2))
\end{equation}
 This expression depends on $A_0$ and the asset volatility $\sigma_A$, both of which are unobservable. However, as shown by Jones et al (cf \cite{hullnelkenwhite}), Ito's lemma can be used to determine the  instantaneous asset volatility from the equity volatility leading to
\begin{equation}
\label{eq:svol}
S_0 \sigma_S \frac{ \partial S }{\partial A} A_0 \sigma_A \end{equation}
Equations (\ref{eq:scall}) and (\ref{eq:svol}) allow the  asset value $A_0$ and the asset volatility $\sigma_A$ to be backed out as functions of the quantities $S_0, \sigma_S, T$ and $L$. Utilizing $A = D + S$, the present value of the debt can be 
thus calculated straightforwardly as
\begin{equation}
D_0 = A_0-S_0
\end{equation}
 where $S_0$ is given by Eq(\ref{eq:scall}).
Now, not that $D_0$ can alternatively be expressed as  
\begin{equation}
D_0 = e^{-yT} D
\end{equation}
Namely as the promised debt payment discounted by some yield $y$.
The difference between this yield $y$  and the risk-free rate $r$ defines the credit spread $s$ on the debt, namely
\begin{equation}
\label{eq:spread}
s = y-r
\end{equation}
This is a very interesting quantity, because it can be shown to be roughly equivalent to premium on a the highly traded credit derivative called a Credit Default Swap (CDS). This is a transaction in which one party  (the protection buyer) pays an annual premium to another party (the protection seller) who basically offers insurance  on the bond of a company. If a default event (which is carefully defined) occurs, the protection seller delivers par to the protection buyer and receives the defaulted bond in exchange. (There are well defined criteria associated with this, somewhat similar to the delivery options in a bond futures contract. Cash settled versions of this contract also exist.)

\section{A non-Gaussian approach}

The driving noise of the asset process in Merton's model is Gaussian, and the related equity process is also a Gaussian process with a constant volatility $\sigma_S$ which can be related to the asset volatility $\sigma_A$ via Eq (\ref{eq:svol}). Thus, the process for stock returns is a standard lognormal one, implying that the stock log-returns are normally distributed across all time scales. Although this is a standard assumption in much of mathematical finance, leading to many interesting and useful results, it is surprisingly far from what one observes empirically. Indeed, real stock returns exhibit fat tails and peaked centers, only slowly converging to a Gaussian distribution as the time scale increases \cite{jpbouchaud,stanley}. One model which generalizes the standard model in a way more consistent with empirical observations has recently been proposed by one of us \cite{prl_borland,qf_borland,qf_borlandbouchaud}. In the spirit of that model, we  generalize the asset price dynamics to follow a non-Gaussian 
statistical feedback process with skew namely
\begin{equation}
dA = \mu_A  dt + \sigma_A A^{\alpha} d\Omega
\end{equation}
\begin{equation}
d\Omega = P^{\frac{1-q}{2}}(\Omega) d\omega
\end{equation}
Here $\omega$ is a standard Brownian noise, $\mu_A$ is the rate of return of the firm  and $\sigma$ is a 
variance parameter.  The parameter $\alpha$ introduces an asymmetry into the model.   $\Omega$ evolves according to a statistical feedback process \cite{pre_borland}, where the probability distribution $P$ evolves according to a nonlinear Fokker-Planck equation 
\begin{equation}
\label{eq:nlfp}
\frac{\partial P}{\partial t}  = \frac{1}{2} \frac{\partial P^{2-q}}{\partial \Omega^2}.
\end{equation}
This diffusion equation maximizes the Tsallis entropy of index $q$ \cite{tsallis,tsallisbukman}. Equation (\ref{eq:nlfp}) can be solved exactly, leading, when the 
initial condition on $P$ is a $P(\Omega,t=0)=\delta(\Omega)$, to a Tsallis 
distribution (equivalent to a Student distribution \cite{andre}): 
\begin{equation}
\label{eq:ptsallis}
P = \frac{1}{Z(t)}\left(1 + (q-1) \beta(t) \Omega^2(t)\right)^{-\frac{1}{q-1}}
\end{equation} 
with 
\begin{equation}
\label{eq:beta}
\beta(t) = c_q^{\frac{1-q}{3-q}}\left((2-q) (3-q) \, t\right)^{-\frac{2}{3-q}}
\end{equation} 
and
\begin{equation}
\label{eq:Z}
Z(t) = \left( (2-q) (3-q) \, c_q t \right)^{\frac{1}{3-q}} 
\end{equation} 
where the $q$-dependent constant $c_q = {\frac{\pi}{q-1}} (\Gamma^2(\frac{1}{q-1}-
\frac{1}{2}))/(\Gamma^2(\frac{1}{q-1}))$.

Eq. (\ref{eq:ptsallis}) recovers a Gaussian in the limit 
$q \rightarrow 1$ while exhibiting power law tails for all $q > 1$.
The index $q$ controls the  feedback into the system. For $q>1$ rare events (small $P$) will give rise to large fluctuations, whereas more common events (larger $P$) yield more moderate fluctuations. 

Pricing options based on this type of model was solved in previous work \cite{qf_borland,qf_borlandbouchaud}. For $q=\alpha=1$, the standard Black-Scholes-Merton model is recovered. For $q=1$ but general $\alpha<1$, the model reduces to the Constant Elasticity of Variance (CEV) model of Cox and Ross (cf \cite{qf_borlandbouchaud}). Closed form solutions for European call options based on this model were found. Evaluating the equity as a call option on the asset process struck at the debt $D$, we thus obtain
\begin{eqnarray}
\label{eq:callskew}
S_0 & = & A_0  \int_{d_1}^{d_2} (1 + (1-\alpha)x(\hat{T}))^{\frac{1}{1-\alpha}} P_q(\Omega_{\hat{T}}) d\Omega_{\hat{T}} 
\nonumber \\
&-& e^{-rT}D\int_{d_1} ^{d_{2}} P_q(\Omega_{\hat{T}}) d\Omega_{\hat{T}}
\end{eqnarray}                
with $x_{\hat{T}}$, $d_1$ and $d_2$ defined in the Appendix.

Just as in the standard Merton framework, it is possible to show that the relationship Eq (\ref{eq:svol}) still holds in the non-Gaussian case. 
Though slightly more complicated, it is still possible to back out $A_0$ and $\sigma_A$ from Eq (\ref{eq:callskew}) and Eq (\ref{eq:svol}).
The debt can then be calculated  as  $D_0 =A_0-S_0$, with $S_0$ as in Eq (\ref{eq:callskew}), and the  credit spread can also be found as in Eq (\ref{eq:spread}), namely
\begin{equation}
\label{eq:spreadqalpha}
s_{q,\alpha} = y-r = -\frac{1}{T} \log( \frac{D_0}{D e^{-rT}})
\end{equation} 
 The main difference is that now the spread is parameterized by $q$ and $\alpha$, 
Because $D_0$ is based on $S_0$ of Eq (\ref{eq:callskew}) allowing us to explore the effects of fat tails ($q >1$) and
skew ($\alpha < 1$).

\begin{figure}[t]
\psfig{file=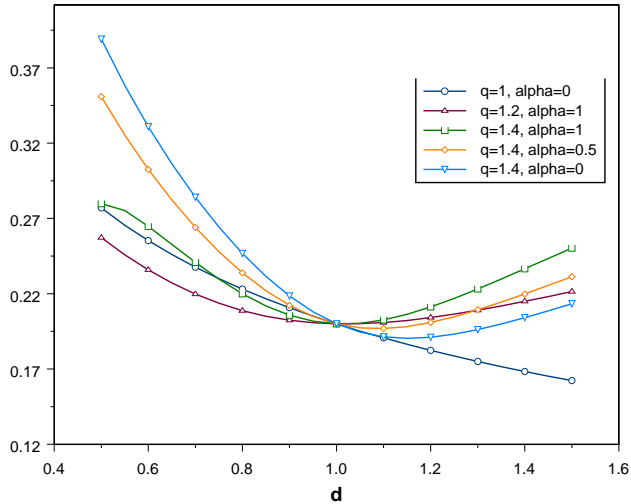,width=4.1 in}
\caption{\footnotesize Implied volatility of the credit spread as a function of $d= D/A_0$ for various values of $q$ (the tail) and $\alpha$ (the skew). }
\end{figure}

\section{Results and Discussion}
Based on the standard Merton model, many practitioners would calculate the fair price of the credit spread from Eq (\ref{eq:spread}). In order to obtain a value of $s$ that matched the market value of the traded CDS, they would have to use a particular value of $\sigma_S$. This implied $\sigma_S$  should in theory 
be the real volatility of the stock. In addition, it should theoretically also be  the correct volatility to price options of the stock. But in practice,  it is already known that it is very difficult to find a single value of $\sigma_S$ that can price all stock options, let alone both stock options and credit spreads. 

In our previous work, we showed that indeed by introducing tails and skew into the underlying stock process,  it was in many cases possible to find 
a single value of $\sigma_S$ that matched well the underlying stock distribution, as well as observed stock option prices. The question we now ask is  whether
the values of $q$ and $ \alpha$ that fit well to the options market
could also well-model observed spreads, or CDS prices. To answer  this question we cite preliminary results  \cite{stanfordstudents} where the parameters  which best fit market spreads across a sample of 54 companies were implied based on 
a least squares fit between the theoretical spread values and market values.
In that analysis, all the observables which enter the spread calculation were taken from reliable data sources as described in \cite{stanfordstudents}.
Certain assumptions were made in order to map the complex capital structure of 
real companies onto the simplified single-debt scenario discussed here. The way in which this was done followed the lines of \cite{hullnelkenwhite}, where a weighted average of all outstanding debt  was used to create a synthetic bond with a single face  value and expiry.
Interestingly, the parameter $q$ was found to be in the range $q= 1.2$ to $q=1.4$, and on average the parameter was $\alpha =  0.3$ with little variation.
These results are quite encouraging, since typical values that are found to fit option markets are very close to  these values. An example is given in \cite{qf_borlandbouchaud} where stock options on MSFT stocks are found to be well fit by $q \approx 1.4$ and $\alpha \approx 0.3$, with $\sigma \approx 0.3$ as well.  (Note that in that example $\sigma_S$ was also implied from the option prices, whereas in the current example, $\sigma_S$ was calculated from historical observations for each entity). 

Finally, we show a figure (Figure 1)  by way of which we want to give a feel for the effect of tails ($q$)  and skew ($\alpha$) on the credit  spread values. We used $r= 4 
\%$, $\sigma_S= 20 \%$, $T = 1$ year and $A_0 = 100$ (in arbitrary monetary units). We varied the ratio $d = D/A_0$ from $0.5$ to $1.5$ and calculated the 
spread according to Eq (\ref{eq:spreadqalpha}) for various combinations of $q$ and $\alpha$, calibrating them by adjusting $\sigma_S$ so that all curves coincide for $d=1$. Then, the corresponding volatility which would have been needed for the standard Merton model ($q=\alpha =1 $)
of Eq (\ref{eq:spread})  to reproduce the same values of the spread were backed out. These implied volatilities were then plotted out as a function of $d$.
Similar in spirit to the implied volatility smiles commonly used to depict the deviations that tails and skew have on option prices relative  to the standard Black-Scholes-Merton model, these curves can give an intuition as to how the spread values  vary relative to the standard log-normal Merton model. 

The standard model corresponds to $q=1$ and $\alpha = 1$. The effect of increasing the tails a bit can be seen in the curves corresponding to $q=1.2$ and $\alpha = 1$: the standard Merton model undervalues leverage ratios greater and less than $1$.  As $q$ increases (see the curve for $q=1.4$ this effect is enhanced. The curve $q=1$ and $\alpha=0$ depicts the effect of skew, and corresponds to a CEV type Merton model. The standard Merton model overvalues 
spreads if there is a debt-to asset ratio greater than 1, and undervalues it otherwise. The effect of both tails and skew can be seen in the other curves. 
The observed behaviour is consistent with the intuition that fat tails correspond to higher  volatilities with respect to a Gaussian model, and increased left skew would correspond to higher volatilities for $d < 1$, and 
relatively lower volatilities for $d >1$.

It is quite easy to visualize where the preliminary empirical results of $q =1 .3 $ and $\alpha = 0.3$  \cite{stanfordstudents}  would lie in this curve. However, further studies must be done to analyze the CDS values systematically as a function of the ratio $d$. In the initial study, the results reported were as average over all the  different companies, each one which had its own  capital structure and leverage ratio. However, encouraged by the fact that the 
parameters which best describe the CDS spreads are in the same ballpark as those
which well-fit stock options as empirical return distributions, it might be worth while to push a little further along the path we have proposed, a project we are currently pursuing.

\section{Appendix}
$x_{\hat{T}}$ is  a function of $\Omega_{\hat{T}}$ given by
\begin{equation}
\label{eq:xhatTPade}
x(\hat T) =  \sigma \Omega_{\hat T} - \frac{\alpha \sigma^2}{2} 
\frac{a(\hat T) + b(\hat T)\Omega_{\hat T} + c(\hat T) \Omega_{\hat T}^2}{1 + d(\hat T) \Omega_{\hat T}}
\end{equation}  
 and
$
\hat{T} = (e^{2(\alpha-1)rT} – 1)/(2(\alpha-1)r).
$
$P_q$ is given by equation Eq (\ref{eq:ptsallis}) evaluated at $t = \hat{T}$.	
The coefficients $a,b,c$ and $d$ are given by 
\begin{eqnarray} \label{eq:padecoeffs}
a &=& g_0(q-1)+\frac{3-q}{2} \gamma\\ \nonumber
b &=&  ad -\eta\tilde{g_1}\\ \nonumber
c&=& (q-1)\frac{\tilde{g_1}}{\eta} d\\ \nonumber
d &=& \frac{g_2(q-1)}{\frac{q-1}{\eta}{\tilde{g}_1} +\eta{g}_1}
\end{eqnarray}
with
$g_0 =\gamma(\hat{T}) \frac{3-q}{2(9-5q)},
g_1 = \gamma(\hat{T}) \frac{(3-q)}{4}$,
$g_2 =\frac{1}{9-5q}$
and
$
\gamma(\hat{T}) =((3-q)(2-q)c_q)^{\frac{q-1}{3-q}}{\hat{T}}^{\frac{2}{3-q}}
$.
The payoff condition $A_T = D$ yields a quadratic equation with the roots
\begin{equation}
d_{1,2} = \frac{N \mp \sqrt{ N^2 - 4MR}}{2M}
\end{equation}
with
\begin{eqnarray}
N &=& -d\frac{ (De^{-rT}/A_0)^{1-\alpha}  - 1}{1-\alpha} +\sigma - b\frac{\alpha \sigma^2}{2}\\
M & = &c \frac{\alpha \sigma^2}{2} -\sigma d \\
R &=&\frac{ (De^{-rT}/A_0)^{1-\alpha} - 1}{1-\alpha}+ a \frac{\alpha \sigma^2}{2}
\end{eqnarray}
These results are as in \cite{qf_borlandbouchaud}.

\end{document}